\documentclass[aps,pra,twocolumn,superscriptaddress,showpacs]{revtex4-1}

\usepackage{amsmath}
\usepackage{graphicx}
\usepackage{epsfig}
\usepackage{dcolumn}
\usepackage{subfigure}
\usepackage{mathrsfs}
\usepackage{multirow}
\usepackage{bbm}
\usepackage{color}
\usepackage[ps2pdf,
bookmarks=true,%                   %%% generate bookmarks ...
bookmarksnumbered=true,%           %%% ... with numbers
hypertexnames=false,%              %%% needed for correct links to figures !!!
]{hyperref}
\newcommand{\bra}[1]{\left< #1 \right|}
\newcommand{\ket}[1]{\left| #1 \right>}
\newcommand{\sbra}[1]{\left( #1 \right|}
\newcommand{\sket}[1]{\left| #1 \right)}

\newcommand{\sbrasket}[2]{\left(\left.\! #1 \right|\! #2 \right)}

\newcommand{\sqrthalf}[0]{\frac{1}{\sqrt{2}}}

\newcommand{\mytitle}{The Impact of Multichannel and Multipole Effects on 
the Cooper Minimum in the High-Harmonics Spectrum of Argon}

\newcommand{\rmpdfinfo}{\special{ps:: userdict /pdfmark /cleartomark load put}}

\definecolor{MyDarkGreen}{rgb}{0,0.6,0}
\definecolor{MyDarkBlue}{rgb}{0,0,0.8}
\definecolor{MyDarkRed}{rgb}{0.6,0,0.3}

\hypersetup{breaklinks=true,
colorlinks=true,
plainpages=true,
linktocpage=true,
linkcolor=MyDarkBlue,
citecolor=MyDarkGreen,
urlcolor=MyDarkRed,
pdfborder={0 0 0},
pdfauthor={Stefan Pabst},
pdftitle={\mytitle},
pdfsubject = {Research Article},
pdfkeywords = {Attosecond, HHG, atomic, argon, strong-field,
               TDCIS, multi-channel},
pdfcreator = {LaTeX with hyperref package},
pdfproducer = {dvipdf}
}

\begin{document} 
%%%%%%%%%%%%%%%%%%%%%%%%%%%%%%%%%%%%
%\setpagewiselinenumbers
%\modulolinenumbers[5]
%\linenumbers

\title{\mytitle}

% Author list 
\author{Stefan Pabst}
\affiliation{Center for Free-Electron Laser Science, DESY, Notkestrasse 85, 22607 Hamburg, Germany}
\affiliation{Department of Physics,University of Hamburg, Jungiusstrasse 9, 20355 Hamburg, Germany}
%\thanks{}

\author{Loren Greenman}
\altaffiliation[Present address: ]{Department of Chemistry, 
Berkeley Center for Quantum Information and Computation, 
University of California, Berkeley, California 94720, USA}
\affiliation{Department of Chemistry and The James Franck Institute, The University of Chicago, Chicago, Illinois 60637, USA}

\author{David A. Mazziotti}
\affiliation{Department of Chemistry and The James Franck Institute, The University of Chicago, Chicago, Illinois 60637, USA}

\author{Robin Santra}
\thanks{Corresponding author}
\email{robin.santra@cfel.de}
\affiliation{Center for Free-Electron Laser Science, DESY, Notkestrasse 85, 22607 Hamburg, Germany}
\affiliation{Department of Physics,University of Hamburg, Jungiusstrasse 9, 20355 Hamburg, Germany}

\date{\today}

\begin{abstract}
We investigate the relevance of multiple-orbital and multipole effects 
during high-harmonic generation (HHG). 
The time-dependent configuration-interaction singles (TDCIS) approach is
used to study the impact of the detailed description of the residual 
electron-ion interaction on the HHG spectrum. 
We find that the shape and position of the Cooper minimum in the HHG spectrum
of argon changes significantly whether or not interchannel interactions are 
taken into account.
The HHG yield can be underestimated by up to 2 orders of magnitude in the 
energy regio of 30-50~eV.
We show that the argument of low ionization probability is not sufficient to
justify ignoring multiple-orbital contributions.
Additionally, we find the HHG yield is sensitive to the nonspherical 
multipole character of the electron-ion interaction.
\end{abstract}
  
\pacs{ 32.80.RM,42.65.Re,31.15.A-}
% 32.80.Rm  Multiphoton ionization and excitation to highly excited states 
% 42.65.Re  Ultrafast processes; 
%           optical pulse generation and pulse compression 
%           (for ultrafast spectroscopy, see 78.47.J-; 
%           for ultrafast magnetization dynamics, see 75.78.Jp)
% 31.15.A-  Ab initio calculations 
% same as in Loren's PRA
%\keywords{}
  
\maketitle
  
  
%%%%%%%%%%%%%%%%%%%%%%%%%%%%%%%%%%%%%%%%%  
\section{Introduction}
\label{sec:intro}
%%%%%%%%%%%%%%%%%%%%%%%%%%%%%%%%%%%%%%%%%%

High-harmonic generation (HHG) is the key physical process underlying the 
generation of single attosecond pulses 
\cite{HeKr-Nature-2001,SaKe-Nature465-2010,SeWa-Nature-2004} 
and attosecond pulse trains 
\cite{PaCh-PRL-1999,PaAg-Science-2001,SiHe-PRL104-2010}, 
which are at the heart of attosecond science \cite{KrIv-RMP81,Bu-Science-2007}. 
In recent years, the rapid progress in HHG has led to applications ranging 
from atomic systems \cite{GiPa-PRL92-2004,UiKr-Nature-2007,LyCo-PRL-2007}
over molecular systems \cite{ItLe-Nature432-2004,BaMa-Science-2006} 
to solid-state systems \cite{CaMu-Nature449-2007}.
HHG has opened a new door to probe structural information
\cite{ItLe-Nature432-2004,SaDu-NJP-2010}
as well electronic and nuclear dynamics \cite{DrKr-Nature-2002,
WoBe-Nature466-2010,GoKr-Nature-2010,WiGo-Science-2011} on fundamental 
time scales.

The mechanism behind HHG is well captured in the three-step model 
\cite{LeCo-PRA-1994}, where in the first step the electron is 
tunnel-ionized by a strong-field laser pulse, in the second step the electron 
is accelerated in the oscillating laser field, and finally in the last step 
the electron recollides with the parent ion and converts its excess energy 
into radiation energy in the extreme ultraviolet range \cite{Co-PRL-1993,
ScKu-PRL-1993}.
The maximum photon energy is given by the cut-off law, $1.32I_p + 3.17 U_p$, 
where $I_p$ is the ionization potential of the system and $U_p$ is the 
ponderomotive potential created by the intense laser field \cite{LeCo-PRA-1994}.
As a result, the heavier noble-gas atoms have lower cut-off energies than the
lighter ones, whereas the HHG yield does increase with the atomic number 
\cite{GoSa-PRL96,VaKa-JPhysB-2011}.
Previous works have shown that the recombination step can be directly related 
to photoionization \cite{LeLi-PRA-2009,LaLi-QM-3edn}, enabling the retrieval 
of the electronic structure of the system 
\cite{ItLe-Nature432-2004,LeLi-PRA-2009,LiLu-JPhysB-2010,MoLi-PRL-2008}.
A strong focus has been, in particular, on molecular systems 
\cite{Vo-PRL-2005,EmSe-NJP-2008,BoSa-NatPhys-2008,ScJo-PRA-2009}.
%However, many aspects such as the role of multiple orbitals
%\cite{SmIv-Nature-2009,McGu-Science-2008,HaSa-NatPhys-2010,WaSm-JPhysB-2010,
%MaHi-PRL104-2010}
%on the HHG spectrum are not completely understood yet. 

The two most common theoretical approaches for describing HHG are the
semiclassical strong-field approximation (SFA) \cite{LeCo-PRA-1994}, which 
has been extended to include Coulomb-interaction corrections 
\cite{LeLi-PRA-2009,KaEh-PRA-1996,SmIv-JPhysB-2007,AbFa-PRA-2009}, 
and the single-active-electron (SAE) approximation 
\cite{KuKr-IntJQuanChem-1991,HiFa-PRA-2011,AwDe-PRA-2008,IvKh-PRA-2009},
where the electron-ion interaction for many-electron systems is described by 
a model potential \cite{HiFa-PRA-2011}.
The SAE approach is computationally more demanding than the SFA approach,
and, therefore, has been limited to atoms and systems like H$_2^+$.
In the literature \cite{LeLi-PRA-2009}, the SAE approximation has often been  
referred to as solving the time-dependent Schr\"odinger equation (TDSE). 
The SAE approach has some limitations.
For example, it ignores contributions from multiple orbitals.
Intensive studies have been performed to understand the impact of 
multiple-orbital contributions in molecular systems, which are essential to 
understand the HHG spectrum and subsequently to extract electronic-structure 
information \cite{McGu-Science-2008,SmIv-Nature-2009}.
Recently, it has been shown that even in atomic systems it is crucial to 
consider multiple-orbital effects \cite{ShVi-NatPhys-2011}.

In this paper, we investigate the importance of multiple-orbital 
(multichannel) contributions and multipole effects in the residual 
electron-ion interaction on the Cooper minimum in the HHG spectrum of argon 
\cite{WoVi-PRL-2009,HiFa-PRA-2011,FaSc-PRA-2011}.
Both aspects are commonly ignored in SFA and SAE calculations.
Multichannel interactions \cite{St-Springer-1980} go beyond the 
independent-particle picture and cannot be captured in the language of SFA and
SAE, whereas multipole effects could, in principle, arise even in a
single-channel model such as SAE.
However, in atomic SAE calculations, it is common to model the electron-ion
interaction by a spherically symmetric potential \cite{HiFa-PRA-2011}. 
The interaction of the liberated electron with the hole state (channel), from
which it originates, is called intrachannel interaction and leads for
large electron-ion distances to the $1/r$ behavior of the Coulomb potential. 
If the liberated electron is influenced by other orbitals the interaction
is called interchannel coupling \cite{St-Springer-1980}. 
The importance of interchannel coupling for HHG has been shown for xenon, where
a clear signature of the giant dipole resonance of the $4d$ subshell
\cite{St-Springer-1980} known from photoionization studies has been directly 
observed in the HHG spectrum of xenon \cite{ShVi-NatPhys-2011}.
The theoretical model we are utilizing to capture these aspects is based
on a time-dependent configuration-interaction singles (TDCIS) approach
\cite{RoSa-PRA74,GrSa-PRA-2010}.
We have demonstrated in previous works that this TDCIS approach is ideal to 
study systematically multichannel effects in situations involving ionization
\cite{RoSa-PRA79-2009,PaSa-PRL106}.

The paper is organized as follows. In Sec. \ref{sec:theory}, we give an 
overview of our theoretical method, which we use to systematically study the
influence of various approximations of the residual electron-ion interaction
on the HHG spectrum.
In Sec. \ref{sec:numerics}, we explain the system parameters used in our 
calculations. The results are discussed in detail in Sec. \ref{sec:results}.
%At the end, we provide a brief summary of our results and explain the impact
%of our results on the common view of HHG and what that means for retrieving
%electronic structure informations.
Atomic units \cite{MoNe-RMP-2008} are employed throughout, unless otherwise 
noted.

%%%%%%%%%%%%%%%%%%%%%%%%%%%%%%%%%%%%%%%%%%
\section{Theoretical Methods}
\label{sec:theory}
%%%%%%%%%%%%%%%%%%%%%%%%%%%%%%%%%%%%%%%%%%  

The time-dependent Schr\"odinger equation of an N-electron system exposed to 
a linearly polarized external electric field is given by
\begin{subequations}
\begin{eqnarray}
  \label{eq:SGL}
  i\frac{\partial}{\partial t} \ket{\Psi(t)}
  &=&
  \hat H(t) \ket{\Psi(t)},
\\
  \label{eq:Hamiltonian}
  \hat H(t)
  &=&
  \hat H_0
  +
  \hat H_1
  -
  E(t) \hat z,
\end{eqnarray}
\end{subequations}
where $\ket{\Psi(t)}$ is the full N-electron wave function and $\hat H(t)$ is 
the exact N-body Hamiltonian, which can be partitioned into three main parts:
(1) $\hat H_0 = \hat F - i \eta \hat W$ is the sum of the time-independent 
Fock operator $\hat F$ and a complex absorbing potential (CAP), where
$\bra{\bf x}\hat W \ket{\bf x'} = [r-r_\text{CAP}]^2\,\Theta(r-r_\text{CAP})\,\delta({\bf x-x'})$ and $\Theta(r)$ is the Heaviside step function.
(2) The electron-electron interactions that cannot be described by the mean-field
potential in $\hat H_0$ are captured by $\hat H_1$ 
($=\hat V_C-\hat V_\text{HF}-E_\text{HF}$; for a detailed description of these
quantities, see Ref. \cite{GrSa-PRA-2010}). 
(3) The term $E(t)\,\hat z$ is the laser-matter interaction in the electric
dipole approximation.
The CAP serves a purely numerical purpose.
It prevents artificial reflections of the ionized photoelectron from the 
radial grid boundary and is located far away from the atom. 
This is controlled by the parameter $r_\text{CAP}$.

Solving numerically the full N-electron system is currently out of reach
without making any approximations to the Hamiltonian or the wave function.
In strong-field processes such as HHG, Eq. (\ref{eq:SGL}) is commonly reduced
to an effective one-electron system, where only one electron of the outermost 
valence shell is allowed to respond to the electric field and all other 
electrons are frozen or completely neglected. 
Here, we take an alternative way, by describing the full N-electron wave 
function and making no approximations to the Hamiltonian.
Specifically, we use the configuration-interaction language, where we assume 
the field-free ground state is well captured by the Hartree-Fock ground state
$\ket{\Phi_0}$.
We consider only singly excited 1-particle-1-hole configurations 
(1p1h-configurations) $\ket{\Phi^a_i}$.
The corresponding TDCIS N-electron wave function reads
\begin{subequations}
\begin{eqnarray}
  \label{eq:wfct_ansatz}
  \ket{\Psi(t)}
  &=&
  \alpha_0(t) \ket{\Phi_0}
  +
  \sum_{i,a} \alpha^a_{i}(t) \ket{\Phi^a_{i}},
\\
  \label{eq:1p1hstate}
  \ket{\Phi^a_{i}}
  &=&
  \sqrthalf \left(
    \hat c^\dagger_{a,\uparrow} \hat c_{i,\uparrow}
    +
    \hat c^\dagger_{a,\downarrow} \hat c_{i,\downarrow}
  \right)
  \ket{\Phi_0},
\end{eqnarray}
\end{subequations}
where $i,j$ and $a,b$ refer to occupied orbitals and unoccupied (virtual) 
orbitals, respectively, in the Hartree-Fock ground state $\ket{\Phi_0}$.
The operators $\hat c^\dagger_{a,\sigma}$ and $\hat c_{a,\sigma}$ create
and  annihilate, respectively, an electron in the orbital $a$ 
with spin $\sigma$.
By restricting our wave function to 1p1h-configurations, the interaction 
captured by $\hat H_1$ is the residual electron-ion interaction.
The equations of motion for the expansion coefficients $\alpha_0(t)$ and 
$\alpha^a_{i}(t)$ read:
\begin{subequations}
\label{eq:eom}
\begin{eqnarray}
  \label{eq:eom_alpha0}
  i\dot\alpha_0(t)
  &=&
  -
  E(t) \sum_{i,a} 
  \sbra{\Phi_0}\hat z\sket{\Phi^a_{i}}
  \alpha^a_{i}(t) 
\\
  \label{eq:eom_alpha_ai}
  i\dot\alpha^a_{i}(t)
  &=&
  (\epsilon_a-\epsilon_i) \alpha^a_{i}(t)
  +
  \sum_{b,j}
  \sbra{\Phi^a_{i}}\hat H_1\sket{\Phi^b_{j}}
  \alpha^b_{j}(t) 
\\&&\nonumber
  - 
  E(t)
  \sbra{\Phi^a_{i}}\hat z\sket{\Phi_0}
  \alpha_0(t)
\\&&\nonumber
  - 
  E(t)
  \sum_{b,j}
  \sbra{\Phi^a_{i}}\hat z\sket{\Phi^b_{j}}
  \alpha^b_{j}(t) 
  ,
\end{eqnarray}
\end{subequations}
where $\epsilon_{p}$ are the orbital energies of the orbitals 
$\ket{\varphi_{p}}$, which are eigenstates of the time-independet Fock operator,
i.e., $\hat H_0 \ket{\varphi_{p}} = \epsilon_{p} \ket{\varphi_{p}}$.
The expression $\sbra{}$ stands for a dual vector with respect to the 
symmetric inner product, i.e., 
$\sbrasket{\varphi_p}{\varphi_q} = \delta_{p,q}$, which differs from the 
Hermitian inner product.
A detailed description of our implementation of the TDCIS method can be found
in Ref. \cite{GrSa-PRA-2010}.

The exact treatment of the residual electron-ion interaction is numerically
very demanding.
In order to be able to treat the full electron-ion interaction, we are 
exploiting as much symmetry as possible. 
We have already used one symmetry with Eq. (\ref{eq:1p1hstate}), i.e., 
the total spin of the system ($S=0$) is conserved.
The second symmetry we are exploiting arises from the restriction to linearly 
polarized pulses and benefits us in two ways.
Firstly, the orbital-angular-momentum projection $m_a$ of the excited electron 
and the orbital-angular-momentum projection $m_i$ of the hole state must be 
the same for each $\ket{\Phi^a_i}$.
Secondly, the coefficients $\alpha^a_i(t)$ are the same whether an 
electron with orbital-angular-momentum projection $m$ or $-m$ is excited.
As a result, only the gerade parity configurations 
$\ket{\Phi^a_i}_g$ need to be considered because ungerade parity configurations 
$\ket{\Phi^a_i}_u$ will not be populated due this symmetry.
The gerade and ungerade parity configurations are defined as 
\begin{eqnarray}
  \label{eq:parity}
  \ket{\Phi^a_i}_{g/u}
  &:=&
  \sqrthalf \left(
  \ket{\Phi^{+a}_{+i}}
  \pm
  \ket{\Phi^{-a}_{-i}}
  \right),
\end{eqnarray}
where the orbital indices $\pm a$ and $\pm i$ stand for triplets of quantum 
numbers $(n,l,\pm m)$ with $n$ being the radial quantum number, $l$ being 
the orbital angular momentum, and $\pm m$ being the orbital-angular-momentum 
projection. 
The configuration $\ket{\Phi^a_i}$  with $m_i=m_a=0$ is a special case and 
has gerade parity.
Since, as already mentioned, for linearly polarized light $\ket{\Phi^a_i}_u$ will
not be populated and only $\ket{\Phi^a_i}_g$ needs to be considered, we drop the 
index $g$ such that $\ket{\Phi^a_i}$ refers to gerade parity configurations 
from now on.  
The matrix elements for the gerade parity configurations as they appear in 
Eqs. (\ref{eq:eom}) are given by
\begin{widetext}
\begin{subequations}
\label{eq:matrix}
\begin{eqnarray}
  \label{eq:mat_dipole_ph}
  \sbra{\Phi^a_i}\hat z\sket{\Phi^b_{j}}
  &=&
  z_{(+a,+b)} \delta_{i,j}
  -
  z_{(+j,+i)} \delta_{a,b}
\\
  \label{eq:mat_dipole_ground}
  \sbra{\Phi_0}\hat z\sket{\Phi^a_{i}}
  &=&
  z_{(+i,+a)}
  \times
  \begin{cases}
    \sqrt{2} &, m_a=m_i=0 \\
    2        &, m_a=m_i\neq 0  \\
    0        &, m_a \neq m_i
  \end{cases}
\\
\label{eq:mat_coulomb}
  \sbra{\Phi^a_{i}}\hat H_1\sket{\Phi^b_{j}}
  &=&
  \left(\Big.
    4 v_{(+a,+j;+i,+b)}
    -
    v_{(+a,+j;+b,+i)}
    -
    v_{(+a,-j;-b;+i)}
  \right)
%\\\nonumber\nonumber  &&
  \times
  \begin{cases}
    1             &, m_i \neq 0 \neq m_j, \\
    \frac{1}{2}   &, m_i   =  0   =  m_j, \\
    \sqrt{2}^{-1} &, \text{otherwise,} 
  \end{cases}
  \qquad
\end{eqnarray}
\end{subequations}
\end{widetext}
where we made use of the symmetries
$z_{(+a,+i)}=z_{(-a,-i)}, v_{+a,+j;+b,+i}= v_{-a,-j;-b,-i}$ and
$ v_{+a,+j;+i,+b} = v_{-a,-j;-i,-b} = v_{+a,-j;+i,-b}$.
In the last equation the relations $m_a=m_i$ and $m_b=m_j$ are used, 
which holds only in the case of linearly polarized light.
The round parentheses in the indices of the matrix elements indicate the 
symmetric inner product mentioned above \cite{GrSa-PRA-2010}.

In the following, we study in detail three scenarios for $\hat H_1$:
(1) no approximation is made and the residual Coulomb interaction is treated
exactly within the CIS configuration space,
(2) only intrachannel interactions 
[$\sbra{\Phi^a_i}\hat H_1\sket{\Phi^b_j}=0$ if $i\neq j$] are considered,
(3) a symmetrized version of the intrachannel interaction is used such that the
angular momentum of the excited electron cannot be changed, thus simulating
a spherically symmetric ion potential.
When only intrachannel interactions are allowed, different orbitals will 
behave almost independently.
Only via the ground-state depopulation can they indirectly influence each 
other. 
The symmetrization in model (3) is done by averaging over all hole states 
within each ($n,l$) subshell such that the excited electron sees only a 
spherically symmetric ion.
The symmetrized matrix elements read
\begin{subequations}
\label{eq:coulomb_symm}
\begin{eqnarray}
  \label{eq:coulomb_symm_a}
  v^\text{symm}_{(a,i;b,i)}
  &:=&
  \frac{1}{2l_i+1}\sum_{m_i} v_{(a,i;b,i)},
\\
  \label{eq:coulomb_symm_b}
  v^\text{symm}_{(a,i;i,b)}
  &:=&
  \frac{1}{2l_i+1}\sum_{m_i} v_{(a,i;i,b)},
\end{eqnarray}
\end{subequations}
where in Eq. (\ref{eq:coulomb_symm_b}) we additionally set $m_a=m_i=m_b$
before we perform the sum.
This step can be justified, since we are using linearly polarized light and
our model can only have 1-particle-1-hole configurations with $m_a=m_i$.
In both cases, one finds that the symmetrized matrix elements are proportional
to $\delta_{l_a,l_b}$ and $\delta_{m_a,m_b}$.

The HHG spectrum, which is calculated via the expectation value of the 
electric dipole moment $\left<z\right>(t)$, reads 
\cite{GoKae-OptExp13,SeSp-PRL-2004} 
\begin{eqnarray}
  \label{eq:spectrum}
  S(\omega)
  &=&
  \frac{1}{20}\frac{1}{3\pi c^3}
  \left|
    \int_{-\infty}^\infty
    \!\! dt \
    \left[\frac{d^2}{dt^2}\left<z\right>(t)\right]
    e^{-i\omega t}
  \right|^2.
\end{eqnarray}
%where the prefactor $1/20$ originates from the fact that we are only 
%looking within 5\% bandwidth.
%
Next to the HHG spectrum, we will focus our discussion also on the hole 
populations $\rho_i(t)$ generated during the HHG process. 
These populations are calculated with the help of the ion density matrix, 
which is described in detail in Ref. \cite{GrSa-PRA-2010}.
The ground-state population is given by $\rho_0(t) = |\alpha_0(t)|^2$.

%%%%%%%%%%%%%%%%%%%%%%%%%%%%%%%%%%%%%%%%%%
\section{Numerical Discussion}
\label{sec:numerics}
%%%%%%%%%%%%%%%%%%%%%%%%%%%%%%%%%%%%%%%%%%  

All presented results were calculated with the XCID package
\footnote[392]{S. Pabst, L. Greenman, and R. Santra - \textsc{XCID} program 
package for multichannel ionization dynamics, DESY, Hamburg, Germany, 2011, 
Rev. 481, with contributions from P. J. Ho}, 
which makes explicit use of the symmetries discussed in Sec. \ref{sec:theory}.
All argon calculations presented in Sec. \ref{sec:results} were done 
for a laser pulse with a peak field strength of $E_\text{max}=0.085$, 
a carrier frequency $\omega=0.057\ (\approx800\,\text{nm})$, and a FWHM pulse 
duration of $\tau=413\ (\approx10\,\text{fs})$.
The classical turning radius of the electron for such a pulse is 
$r_\text{HHG}=E_\text{max}/\omega^2 \approx 26$.
As described in Ref. \cite{GrSa-PRA-2010}, we use a non-uniform grid with the 
mapping function
\begin{eqnarray}
  r(x)
  &=&
  r_\text{max} \frac{\zeta}{2} \  \frac{1+x}{1-x+\zeta} 
  \quad, x\in[-1,1].
\end{eqnarray}
All calculations were done with a radial grid radius 
$r_\text{max}=120$, 480 radial grid points, and mapping parameter $\zeta=1.0$.
The CAP starts at a radius $r_\text{CAP}=100$ and has a strength $\eta=0.01$.
The maximum angular momentum employed was $l_\text{max}=80$.
Furthermore, we find that for excited electrons with an orbital angular 
momentum $l>6$ the multipole terms with $L_c>0$ are negligibly small and the 
dominant $\hat H_1$ contribution comes from the monopole term.
It is, therefore, a good approximation to consider only the monopole term
of $\hat H_1$ when any orbital angular momentum of the involved orbtials
is larger than 6 [for details see Ref. \cite{GrSa-PRA-2010}].

%%%%%%%%%%%%%%%%%%%%%%%%%%%%%%%%%%%%%%%%%%
\section{Results}
\label{sec:results}
\begin{figure}[ht!]
\begin{center}
  \rmpdfinfo
  \includegraphics[clip,width=.7\linewidth]{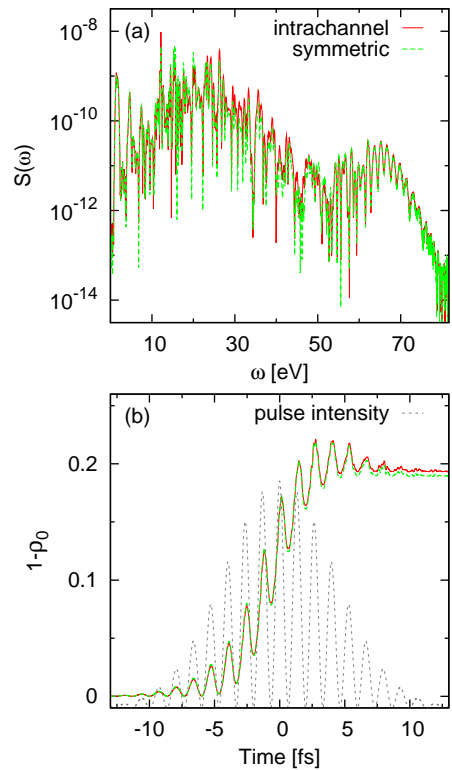}
  \caption{(color online) The HHG spectra (a) and the  depopulations of the
    ground state (b) of argon are compared for the intrachannel approximation
    (red solid line) and for the symmetrized intrachannel approximation 
    (green dashed line). 
    In both cases only $3p_0$ (single-channel) is active. 
    The intensity profile of the pulse is shown in (b). The pulse parameters 
    are: $E_\text{max}=0.085$, $\omega=0.057\ 
    (\approx800\,\text{nm}),~\text{and}~\tau=413\ (\approx10\,\text{fs})$.
  }
  \label{fig:3p0}
\end{center}
\end{figure}

%%%%%%%%%%%%%%%%%%%%%%%%%%%%%%%%%%%%%%%%%%  
We begin our discussion with the single-channel model by allowing only 
the $3p_0$ orbital to be active.
In Fig. \ref{fig:3p0}, we compare the HHG spectra and the depopulations of the
ground state of argon for a spherically symmetric electron-ion interaction 
(labeled symmetric) with the exact electron-ion interaction (labeled 
intrachannel). 
Note, interchannel contributions do not exist in a single-channel model.
The spherically symmetric $\hat H_1$ has no tensorial multipole moments besides
a monopole term, since the angular momentum of the electron cannot be changed. 
In the intrachannel and interchannel models all multipole contributions are 
included in $\hat H_1$.
The depopulations [shown in Fig. \ref{fig:3p0} (b)] show only small deviations
during and after the pulse.
The final depopulation probabilities are almost identical. 
Similarly, the HHG spectra [see Fig. \ref{fig:3p0} (a)] show only small differences in the energy region of 30-50~eV, where also the photoionization cross sections (not shown) differ by up to $30\%$ from each other.
The Cooper minimum in the HHG spectra can be reproduced and lies 
between 40-50~eV.
The low curvature of the shape of the Cooper minimum prevents a more precise localization of the minimum.

Due to the costly treatment of the residual electron-ion interaction it is 
common to reduce an HHG calculation to a one-electron calculation, where the 
electron moves in a local, spherically symmetric model potential, which 
describes the correct behavior for short and long distances of the electron-ion
interaction and reproduces the ionization potential.
This is the SAE approach, which is in spirit very close to our single-channel
model with a spherically symmetric $\hat H_1$ (green dashed line in Fig. 
\ref{fig:3p0}).
However, there exists one major difference to typical SAE calculations.
In the SAE approach one only describes one electron, which can move
freely everywhere on the pseudopotential surface, and does not fulfill
the Pauli principle, meaning there is no mechanism in this approach that 
can prevent the electron to move into orbitals that are already occupied 
by the $N-1$ frozen electrons. 
Enforcing the Pauli principle is critical for the one-electron reduced density
matrix to be N-representable, that is, to represent a realistic N-electron 
system \cite{RoMa-JCP-2010,Ma-PRL-2011}.
In our theory,  we describe always the entire $N$-electron wave function and 
due to the anticommutator relation of the creation and annihilation operators 
in Eq. (\ref{eq:1p1hstate}) the Pauli principle is ensured at all times. 
Recent works for molecular systems have pointed out the importance of the 
Pauli principle particularly for tomographic purposes 
\cite{SaGo-PRL-2006,PaVi-PRL-2006}.

\begin{figure*}[ht!]
\begin{center}
  \rmpdfinfo
  \includegraphics[clip,width=\linewidth]{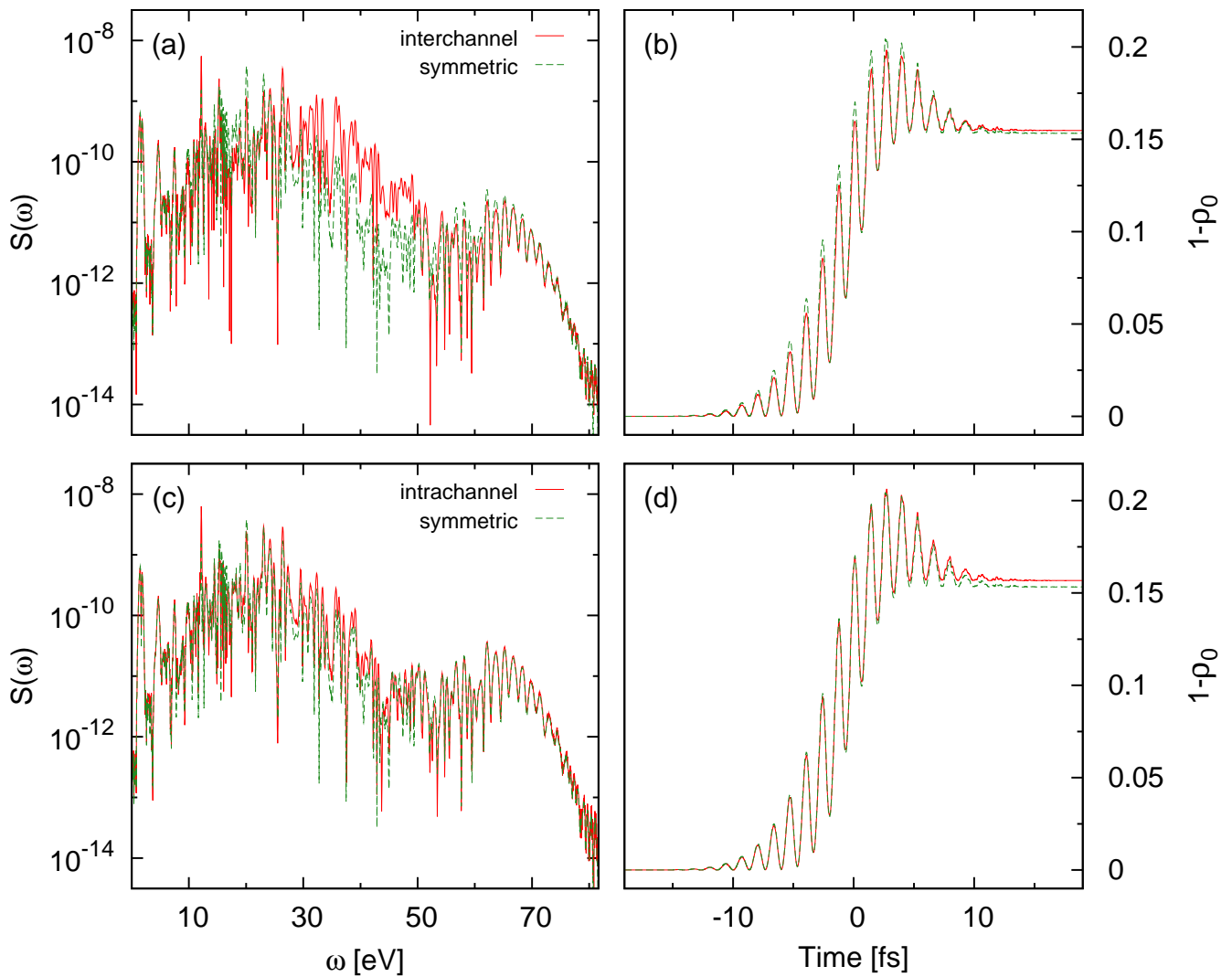}
  \caption{(color online) 
    The HHG spectra (a,c) and the depopulations of the ground state (b,d) of 
    argon are shown for $3s$ and all $3p$ orbitals active. 
    In (a,b) the interchannel and in (c,d) the intrachannel 
    approximation is compared with the symmetrized intrachannel approximation. 
    The pulse parameters are the same as in Fig. \ref{fig:3p0}.
  }
  \label{fig:interaction}
\end{center}
\end{figure*}

\begin{figure*}[ht!]
\begin{center}
  \rmpdfinfo
  \includegraphics[clip,width=\linewidth]{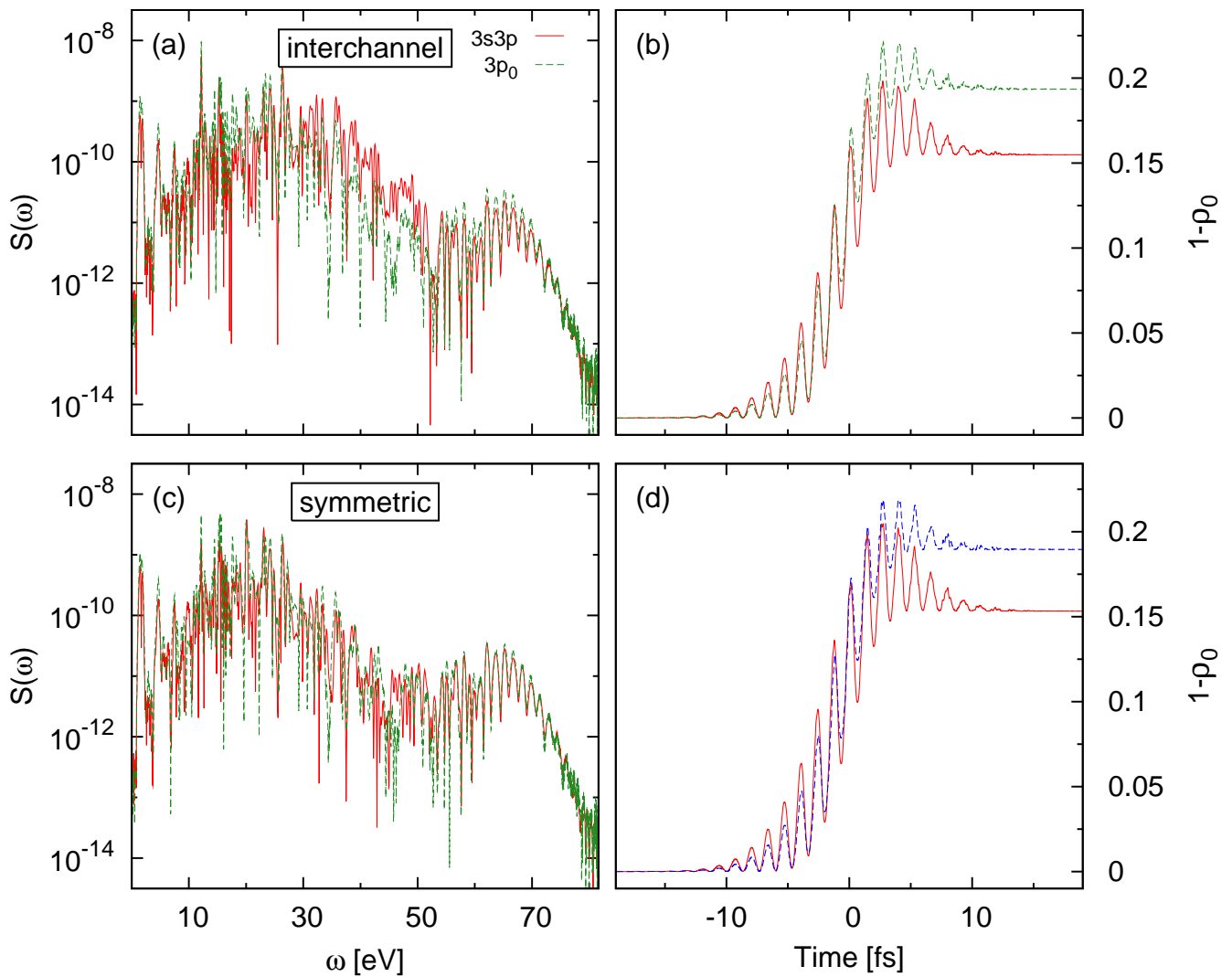}
  \caption{(color online) 
    The HHG spectra (a,c) and the depopulations of the ground state (b,d) 
    of argon are shown. 
    The single-channel (only $3p_0$ active) calculation is compared 
    with the multichannel calculation ($3s$ and all $3p$ active).
    No approximation (a,b) and the symmetrized intrachannel approximation (c,d)
    are made to the electron-ion interaction.
    The pulse parameters are the same as in Fig. \ref{fig:3p0}.
  }
  \label{fig:multichannel}
\end{center}
\end{figure*}

\begin{figure}[ht!]
\begin{center}
  \rmpdfinfo
  \includegraphics[clip,width=.7\linewidth]{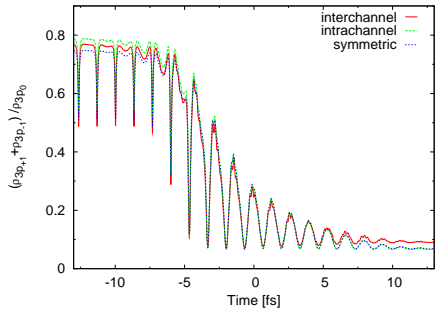}
  \caption{(color online) 
    The ratio of the hole populations 
    $\rho_{3p_1}+\rho_{3p_{-1}}$ and $\rho_{3p_0}$ is shown for different 
    approximations to the electron-ion interaction. 
    The pulse parameters are the same as in Fig. \ref{fig:3p0}.
  }
  \label{fig:ratio}
\end{center}
\end{figure}

Now we consider the impact of the approximation of the residual 
electron-ion interaction in the multichannel scenario, where 
we allow also $3p_{\pm 1}$ electrons and the $3s$ electrons to get ionized.
All differences seen in Figs. \ref{fig:interaction}, \ref{fig:multichannel}, 
and \ref{fig:ratio} originate solely from physics within the $3p$ manifold.
The impact of the $3s$ orbital is rather small on the HHG spectra as well 
as on the depopulation of the ground state. 
That is not a surprise due to the high $3s$ ionization potential, which is 
around 18~eV higher than the ionization potential of the $3p$ orbitals. 
The hole population of $3s$ (not shown) is over 100 times smaller than the 
hole populations in the $3p$ shell.

In Fig. \ref{fig:interaction} the HHG spectra and depopulations are shown for
different approximations of $\hat H_1$.
The simplification to a spherically symmetric potential 
[see Fig. \ref{fig:interaction} (a,b)] underestimates the HHG spectrum by up
to two order of magnitude in the energy region of 30-50~eV.
In addition, the shape of the Cooper minimum has now drastically changed.
The position of the Cooper minimum is much more clearly defined in the interchannel
case and lies slightly above 50~eV as found in recent experiments 
\cite{WoVi-PRL-2009,HiFa-PRA-2011,FaSc-PRA-2011}.
For the symmetrized intrachannel approximation, the Cooper minimum lies between 
40-50~eV, similar to the single-channel results.
The depopulation dynamics is not affected by the approximation of
the electron-ion interaction.
In Fig. \ref{fig:interaction}(c,d) the results of the intrachannel
calculation are compared with the results obtained from the symmetrized 
intrachannel approximation. 
The relative differences between these two models never exceed a factor larger
than 2 and are confined to the energy region of 30-50~eV. 
The origin of these differences is the lack of multipole effects in the 
symmetrized intrachannel model.
In comparison to the interchannel results, the multipole effects are much
smaller than the interchannel effects seen in Fig. \ref{fig:interaction}(a).

The photon energy range of 30-50~eV corresponds to a recollision electron energy 
range of 15-35~eV and a de Broglie wavelength of 2-3~\AA. 
It seems that electrons with these wavelengths are most sensitive to the exact 
residual ion-electron interaction and, therefore, simplifications of the 
interaction become most evident in the corresponding photon energy regime. 
In the same energy region the photoionization cross section is 
most sensitive to the approximation made to the electron-ion interaction 
\cite{St-Springer-1980}.
Our calculations (not shown) confirm that the differences in the cross sections 
between the intrachannel and symmetrized intrachannel model are quite small, 
where as the differences to the interchannel model are up to
one order of magnitude larger (and can reach values up to 20~Mb). 
The fact that the photoionization cross sections and the HHG spectra behave similarly
(for different approximations to the electron-ion interaction) supports the 
picture that HHG has a close connection to photoionization.
In the limit that the electron de Broglie wavelength is much longer or shorter 
than the characteristic length scale of the residual electron-ion interaction,
the results do not depend on the detailed structure of electron-ion interaction.
This may explain why the HHG spectrum does not alter significantly for photon 
energies smaller than 20~eV and photon energies close to the cut-off region.

We have seen that approximations to the electron-ion interaction can cause
differences in the HHG yield, particularly, when multiple orbitals are 
considered. 
Not all electron-ion approximations make the HHG spectrum sensitive to 
whether a single orbital or multiple orbitals are active.
In Fig. \ref{fig:multichannel} we compare the HHG spectrum and the 
depopulation of the ground state for a single-channel (only $3p_0$ active) and
for a multichannel (all $3p$ and $3s$ are active) calculations.
Figures \ref{fig:multichannel}(a-b) are calculated with the
exact $\hat H_1$ term including interchannel and multipole effects 
in the residual electron-ion interaction.
The HHG signal strength for the single-channel calculation is strongly reduced
in the spectral range of 30-50~eV, whereas the signal is slightly enhanced in
the cut-off region.
When the symmetrized intrachannel approximation is made 
[see Figs. \ref{fig:multichannel}(c-d)],
the HHG spectra are almost identical whether single-channel or 
multichannel calculations are performed.
This stands in contrast to the interchannel results, where the interchannel 
coupling causes strong differences in the HHG yield. 
The depopulation is overestimated by $\approx 25\%$ regardless of the 
approximation made to the electron-ion interaction. 

We have seen the strong differences in the HHG spectra and in the
depopulations depending on whether only $3p_0$ or all $3p$ electrons
are active (the contributions from $3s$ are negligible small). 
Does that also mean the ionization probabilities of $3p_{\pm 1}$ are 
comparable with $3p_0$ ?
The ratio $(\rho_{3p_1}+\rho_{3p_{-1}})/\rho_{3p_0}$ is shown in Fig. 
\ref{fig:ratio} for the differenet models.
Note that for linearly polarized light $\rho_{3p_{+1}}=\rho_{3p_{-1}}$.
Before the pulse, all ratios are close to 0.8 and drop to $\approx 0.1$ after
the pulse.
Both intrachannel models lead even to the same final ratio.
The small ratios after the pulse show that, at least ultimately, mainly the 
$3p_0$ orbital gets ionized, which is at the heart of the SAE approximation.  
The oscillations in the ratios during the pulse are in phase with the 
oscillations in the electric field.
They are a direct consequence of the projection of the wave function onto the 
field-free states in the presence of the laser field.
In contrast to the HHG spectra (see Fig. \ref{fig:interaction}), the 
effects of the intrachannel or symmetric approximation on the population dynamics 
are quite small.

%%%%%%%%%%%%%%%%%%%%%%%%%%%%%%%%%%%%%%%%%%
\section{Conclusion}
\label{sec:conclusion}
%%%%%%%%%%%%%%%%%%%%%%%%%%%%%%%%%%%%%%%%%%
We have described the HHG process with a many-body approach, namely TDCIS, 
where we describe the entire N-electron wave function.
This allows us to fulfill the Pauli principle at all times.
Our results show that multichannel effects in the residual electron-ion interaction, 
which is a combination of the bare nuclear potential and the electron-electron 
interaction for many-electron systems, have a significant influence on the HHG 
spectrum.
They cannot generally be neglected for atoms and specifically not for 
molecules as recent experiments have shown 
\cite{McGu-Science-2008,SmIv-Nature-2009}.
We have demonstrated that orbitals, despite their relatively low ionization 
probability by the end of the pulse, can lead to surprisingly large 
modifications of up to 2 orders of magnitude in the HHG spectrum
(especially in the energy region of 30-50~eV).
While we confirm that after the end of the pulse, the populations of the 
$3p_{\pm1}$ orbitals are relatively small, their contributions during the pulse
are not small and have indirectly through interchannel coupling a significant 
impact on the HHG yield.

We saw that neglegting interchannel interactions lead to large changes in the 
HHG yield.
Multipole effects influence the spectra but not as dramatically as
interchannel effects do. 
All deviations in the HHG yield are in the 30-50~eV energy region, which 
corresponds to a de Broglie wavelength of the recolliding electron between
2-3~\AA.
This coincides with the characteristic length scale on which the electron-ion
interaction goes over into a pure long-range $1/r$ potential.
In contrast to the large disagreement in the HHG spectra between the 
single-channel and multi-channel calculations including interchannel interactions
we found that by using the symmetric interaction the HHG spectra look quite the 
same whether or not a single or multiple channels participate in the HHG process. 
This comparison directly shows that the population of an orbital does not map 
directly to its importance in the HHG mechanism.
 
All these observations demonstrate that many-body effects enter in the HHG 
spectrum and need to be understood in order to successfully use them for 
tomographic imaging \cite{ItLe-Nature432-2004}.
The time-dependent configuration-interaction approach provides a clear pathway 
how these and higher-order effects can be taken into account.
Recent works \cite{SuBr-PRL-2009} have suggested that multielectron 
excitations are not a dominant factor.
All essential multielectron effects can be captured by single-electron 
excitations including interchannel interactions. 
This makes the TDCIS approach perfectly suited for studying many-body effects 
in HHG.

\acknowledgments
This work has been supported by the Deutsche Forschungsgemeinschaft (DFG) 
under grant No. SFB 925/A5.
We thank Sang-Kil Son for helpful discussions.

% use bibtex
%\bibliographystyle{plainnat}
\bibliography{amo,books,solidstate}

\end{document}